\documentclass[10pt, conference, compsocconf]{IEEEtran}
\usepackage{booktabs} 
\usepackage{tikz}
\usepackage{tkz-graph}
\usepackage{tkz-berge}
\usepackage{pgfplots}
\usepackage[ruled,vlined,linesnumbered,algo2e,resetcount]{algorithm2e}
\usepackage{algorithmic}
\usepackage{cite}
\usepackage{caption}
\usepackage{amssymb,amsmath,amsthm}
\newtheorem{theorem}{Theorem}

\newtheorem{Lemma}{Lemma}
\usepackage{chngcntr}
\usepackage{graphicx}
\usepackage{epstopdf}
\usepackage[para]{threeparttable}
\usepackage{booktabs, caption, makecell}
\usepackage{amsmath}
\usepackage{cancel}
\usetikzlibrary{shadows}
\usepackage{float}
\usepackage{balance}
\usepackage{cancel} 
\usepackage{url}  
\usepackage{multirow}
\let\oldnl\nl
\newcommand{\nonl}{\renewcommand{\nl}{\let\nl\oldnl}}%
\usepackage{tabu}
        
\tikzstyle{nodestyle}=[draw,circle, draw=black!100]
\tikzstyle{selectnodestyle}=[draw,circle,circular drop shadow, draw=black!100,fill=red!40]
\tikzstyle{selectedgestyle}=[draw,shadow]
\SetKwProg{Fn}{Function}{}{}

\begin{document}

\title{The estimation of bias and variance in clustering coefficient streaming algorithms}

	\author{\IEEEauthorblockN{Roohollah Etemadi, Jianguo Lu}
		\IEEEauthorblockA{School of Computer Science, University of Windsor\\
			Windsor, ON, Canada\\
			{etemadir, jlu}@uwindsor.ca}		
	}

\maketitle


\begin{abstract}
Clustering coefficient is one of the most important metrics to understand the complex structure of networks. 
This paper addresses the estimation of clustering coefficient in network streams. There have been substantial work in this area,  most of conducting empirical comparisons of various algorithms. The variance and the bias of the estimators have not been quantified.  Starting with a simple yet powerful streaming algorithm, we derived the variance and bias for the estimator, and the estimators for the variances and bias. More importantly, we simplify the estimators so that it can be used in practice. The variance and bias estimators are verified extensively on 49 real networks. 
           
\end{abstract}




\IEEEpeerreviewmaketitle

\section{Introduction}
 The amount of data in digital world is growing faster than before in the age of the Internet. This deluge of data results in massive networks with the size of billions. Recently analyzing such real-world networks has captivated a great attention among practitioners and scholars. Clustering coefficient (hereafter $\mathcal{C}$) is one of the most important metrics to analyze such networks.  It has been used in many applications including graph clustering, community detection, spam detection, link prediction, wireless and ad-hoc networks analysis, microarray data and DNA sequence analysis, word-learning, risk analysis, etc. Computing $\mathcal{C}$ on large networks is an intensive task, i.e. the time complexity of the state-of-the-art method is $\Theta(N^{1.41})$, here $N$ is the number of nodes in the network \cite{latapy2008main}. Thus, sampling-based algorithms are indispensable. 
 
 This paper focuses on the streaming algorithms where data arrive sequentially in an arbitrary order. A number of techniques have been proposed to estimate $\mathcal{C}$ in such a streaming model \cite{ahmed2014graph,ahmed2017sampling,jha:2013:SES,jha2015space}. The aim is to provide an accurate estimation using limited memory budgets over single or multi passes of the stream of edges of networks. 
 
 Despite extensive work in this area, 
 there is a lack of formal analyses of the algorithms, in particular the lack of estimators for the variance and bias of the estimation algorithms. Most algorithms are compared empirically, thus their performances are often data dependent. Some algorithms do discuss the bias and variance problems, but they all fail to give the exact formulas. E.g., \cite{ahmed2014graph}  \cite{ahmed2017sampling} gives a variance that is generic for every problem; \cite{ahmed2014graph,ahmed2017sampling,jha:2013:SES,jha2015space} mentioned about the existence of the bias. 
  
In addition to the bias and variance formulas, a more important issue is their estimation without the global data. In order to have the confidence interval for an estimate, we need to know the variance and bias during the sampling and estimation process, not the variance and bias that is derived from the knowledge of the global data. I.e., we need to estimate the variance and bias.  
 
This paper derived the variance and bias so that algorithms can be compared analytically. The formulas are along and tedious,  hard to be applied in practice. A more interesting result of the paper is the simplification of the formulas so that it can be applied easily. Based on the assumption that the data is big, we show that the RSE (Relative standard error, the normalized square root of variance) can be estimated by the reciprocal of the square-root of the triangles observed in samples. In other words, the accuracy grows at the rate of $\sqrt{\Delta_g}$, where $\Delta_g$ is the observed sample closed wedges in the streaming process.  

The simplified result is especially important in the era of big data. When networks are small, we can hardly predict the behaviour of the sampling algorithms. However, when the data is big, the performance can be characterized simply by the closed wedges observed in the process. This result is also supported by our extensive experiments on 49 real network of different size and structure. 
 
 
 
The second contribution is the correction of the bias of the traditional estimator. This is an extension of the work from direct sampling \cite{etemadi2017cbias} to streaming algorithms. The existence of the bias in streaming algorithms was also observed in \cite{ahmed2017sampling}, but it is not quantified, thus not corrected. We proved that the bias depends on the structure of the sample. 

 \section{Background and related work}
\subsection{Notations}Suppose $G(\mathcal{V}, \mathcal{E})$ be a simple undirected graph, where $\mathcal{V}$ is the set of nodes, and $\mathcal{E}$ the set of edges.  For simplicity, we assume that the graph is not a multi-graph, and does not have self-loops. Let $N=|\mathcal{V}|$, $M=|\mathcal{E}|$.   Let $1,2, \dots, M$ be the labels of the edges in $\mathcal{E}$ according to their occurrence order in the stream. E.g. edge $M$ is the last edge in the stream. 
A \textit{wedge} $\mathcal{W}$ is a path $(u,v,w)$ of length two, where nodes $u, v, w \in \mathcal{V}$, and edges $(u,v) \in \mathcal{E}$ and $(v, w) \in \mathcal{E}$. A wedge $\mathcal{W}$ is closed if $(u, w)\in \mathcal{E}$. Otherwise it is open. A closed wedge $\mathcal{W}$ is also called a triangle. Note that each triangle has three (closed) wedges. Let $\Delta$ denote the number of closed wedges and $\Lambda$ the count of wedges in $G$. 

\subsection{Related work}A number of sampling-based methods have been proposed in recent years to estimated clustering coefficient in networks. A straightforward method is \textit{wedge sampling} \cite{schank2004approximating}. It selects wedges uniformly at random and the fraction of closed ones is used as an estimation for $\mathcal{C}$. To sample a random wedge, two passes over an edge stream are required in a streaming model. Furthermore, one additional pass over edge stream is needed to check the closeness of the random wedge \cite{jowhari2005new} \cite{buriol2006counting}. 
Therefore, sampling a random wedge from a large graph is an intensive task. A scalable technique to estimate $\mathcal{C}$ is \textit{edge sampling} \cite{ahmed2014graph,ahmed2017sampling,jha:2013:SES}. It samples some edges uniformly at random and create a subgraph $g$. Then, the number of wedges and closed ones in $g$ are used to estimate $\mathcal{C}$.

Estimating $\mathcal{C}$ using edge-based sampling methods is biased. It is also noticed in \cite{ahmed2014graph}\cite{ahmed2017sampling}\cite{jha2015space}. Recently, authors in \cite{etemadi2017cbias} quantify the bias in none-streaming model when edges of original graph are randomly accessible. However, to the best of our knowledge, there is no study to quantify the bias in the streaming model.     


 		
 \begin{algorithm2e}[t]
 	\DontPrintSemicolon
 	\SetAlgoLined 
 	\small	
 	\KwIn{$p$  }
 	\KwOut{$\widehat{\Delta}$, RSE($\widehat{\Delta}$)}
 	\SetKwFunction{funu}{Update$\Lambda_c\&\Psi_g\&\Omega_g$}
 	\SetKwFunction{fund}{Update$\Delta_g\&\Phi_g$}
 	\Begin{	
 		$\Delta_g=0$, $\Lambda_g=0$, $g=\{\phi\}$.	\\	
 		\While{new edge e}{
 			Add $e$ into $g$ with probability $p$.\\
 			\ForEach{wedge $w$ formed using $e$ and edges in $g$}{$\Lambda_g+=1.$}
 			\ForEach{wedge $w\in g$ closed by $e$}{$\Delta_g+=1.$} 
 			
 		}
 		$\widehat{\mathcal{C}}=3\Delta_g(p\Lambda_g)^{-1}$.\\
 		$\widehat{RSE}(\widehat{\mathcal{C}}$)$\approx \Delta_g^{-1/2}$.	
 	} 
 	
 	\caption{ Naive edge sampling (NES)} \label{Alg1}
 \end{algorithm2e} 		
\subsection{The Algorithm}
  In order to discuss the variance and bias problem rigorously, we give a simple edge-based streaming algorithm (hereafter we call it NES--Naive Edge Streaming) as described in Alg. \ref{Alg1}. It is adapted from a direct sampling algorithm \cite{etemadi2017cbias} to the stream model.   It is also a special case of \cite{ahmed2017sampling} where all edges are sampled with the same priority.
  
  In NES, edges arrive in an arbitrary sequence over an edge stream of the original graph $G$. Over the stream, NES samples some edges uniformly at random with an equal probability $p$,  and adds them into the subgraph $g$.  Once a new edge $e$ arrives, the number of wedges closed by $e$ in $g$ is counted. 
 Let $\Delta_g$ denote the  number of such closed wedges observed in the stream.  At the same time, the number of wedges formed using $e$ and edges in $g$ is enumerated. Let $\Lambda_g$ be the total number of such wedges. 
 Then, $\mathcal{C}$ is estimated using $\Delta_g$ and $\Lambda_g$. A key difference with \cite{etemadi2017cbias} is that $\Delta_g$ and $\Lambda_g$ are counted along the streaming process, not after the sampling.  

 Given a subgraph $g$ of $G$, for every wedge $(u,v,w)$ in $g$,  we check whether or not $(u,w)$ is in the rest of the stream. For every closed wedge in $G$, if its two edges are sampled in $g$, the probability of observing the third edge in the rest of the stream is $\frac{1}{3}$. Thus, the probability of identifying a closed wedges by NES is $\frac{1}{3}p^2$. 
 Suppose $\delta_i$ be an indicator for the $i^{th}$ closed wedge in $G$. Variable $\delta_i$ is 1  when two edges of the $i^{th}$ closed wedge are sampled and the third edge is observed in the rest of the stream;  otherwise it is 0. 
 Recall that $\Delta_g$ is the number of closed wedges identified by NES over a single pass on an edge stream of $G$. The expectation of $\Delta_g$ is 
 $
 \mathbb{E}(\Delta_g)= \mathbb{E}(\sum_{i=1}^{\Delta} \delta_i) 
 =  \sum_{i=1}^{\Delta} \mathbb{E}(\delta_i) 
 =  \sum_{i=1}^{\Delta} \frac{1}{3}p^2 
 = \frac{1}{3} p^2 \Delta$. Thus, an unbiased estimator for $\Delta$ is given by
$ \widehat{\Delta}= \frac{3\Delta_g}{p^2}$. Similar estimators for $\Delta$ have also been proposed in \cite{lim2015mascot,etemadi2016triangles,DeStefani2016TRIEST2016graph}. 
 
Next, we give an unbiased estimator for the number of wedges in $G$. To sample a wedge by NES, one of its two edges needs to be added into $g$ with probability $p$, and its second edge is required to be observed in the rest of the stream. Thus, the probability of identifying a wedge based on $g$ is $p$. 
 Suppose $\lambda_i$ be an indicator for the $i^{th}$ wedge in the input graph. Clearly, $\lambda_i$ is 1  when its two edges are observed;  otherwise it is 0. Recall that $\Lambda_g$ is the number of wedges identified based on $g$ by NES. Its expectation is
$\mathbb{E}(\Lambda_g)= \mathbb{E}(\sum_{i=1}^{\Lambda} \lambda_i) 
=  \sum_{i=1}^{\Lambda} \mathbb{E}(\lambda_i)=\sum_{i=1}^{\Lambda} p=p\Lambda$. Thus, an unbiased estimation for $\Lambda$ is given by
$\widehat{\Lambda}= \frac{\Lambda_g}{p}. 
$

Now we can use the unbiased estimators for $\Delta$ and $\Lambda$ to estimate $\mathcal{C}$.
Although both $\widehat{\Delta}$ and $\widehat{\Lambda}$ are unbiased, the following estimator is \textbf{biased} as we will correct it later in this paper.
\begin{equation}
	\widehat{\mathcal{C}}= \frac{\widehat{\Delta}}{\widehat{\Lambda}}
	= \frac{3p\Delta_g}{p^{2}\Lambda_g}=\frac{3\Delta_g}{p\Lambda_g}.   \label{est}
\end{equation}
 
\section{Estimator of the variance}
We derive the variance of the estimator using the Delta method. Applying $var$ on Eq. \ref{est}, we get
\begin{align}
var(\widehat{\mathcal{C}})&=var\bigg(\frac{3\;\Delta_g}{p\;\Lambda_g}\bigg)
=\frac{9}{p^2}var\bigg(\frac{\Delta_g}{\Lambda_g}\bigg). \label{var_c1}
\end{align}
Applying Taylor expansion in the neighbourhood of $(a,b)$, we have: 
\begin{align}
var\biggl(\frac{\Delta_g}{\Lambda_g}\biggr)\approx&\frac{1}{b^2}var(\Delta_g)+\frac{a^2}{b^4}var(\Lambda_g)
-\frac{2a}{b^3}cov(\Delta_g,\Lambda_g).   \notag
\end{align}
Let $a=\mathbb{E}(\Delta_g)$ and $b=\mathbb{E}(\Lambda_g)$, and using the fact that  $\mathbb{E}(\Delta_g)=\frac{1}{3}\Delta p^2$, and $\mathbb{E}(\Lambda_g)=\Lambda p$, we obtain the variance and present it in the form of relative standard error (RSE=$\sqrt{var}/\mathcal{C}$) as follows:

\begin{align}
RSE(\widehat{\mathcal{C}})&\approx \biggl[\frac{9\;var(\Delta_g)}{\Delta^2p^4}+\frac{var(\Lambda_g)} {\Lambda^2p^2}
-\frac{6\;cov(\Delta_g,\Lambda_g)}{\Delta\Lambda p^3}\biggr]^{-1}. \notag 
\end{align}
The RSE depends on the variance of $\Delta_g$ and $\Lambda_g$, and the covariance between them. 
Note that this is where \cite{ahmed2014graph,ahmed2017sampling} stops.  We continue the derivation of the variances and covariances in Appendix. When the networks are large, $p$ is a very small value.  Hence, we can assume that $1-p\approx1-p^2 \approx 1$, and the results of the Lemmas \ref{lem:varDelta}, \ref{lem:varLambda}, and \ref{Lem:cov} are simplified as follows: 
\begin{align}
&var(\Delta_g)\approx\frac{1}{3}\Delta\;p^2+\frac{16}{15}\Phi\;p^3, \label{approxvarDelta}\\
&var(\Lambda_g)\approx  \frac{2}{3}\Psi\;p, \label{approxvarLambda}\\
&cov(\Delta_g,\Lambda_g)\approx  \frac{5}{12}\Omega'\;p^2. \label{approxCov}
\end{align}
Here $\Phi$ is the number of pairs of dependent triangles, $\Psi$ is the number of pairs of shared wedges, and $\Omega'$ is the number of pairs of wedges and triangles with one common edge in the original graph $G$.
Replace Eq.s \ref{approxvarDelta}, Eq. \ref{approxvarLambda}, and Eq. \ref{approxCov} in the RSE above and after some math simplifications we obtain the RSE as:


\small \begin{align}
RSE(\widehat{\mathcal{C}})&\approx \biggl[\frac{3}{\Delta p^2}+\frac{48\Phi p^3}{5\Delta^2p^4}+ \frac{2\Psi\;p}{3\Lambda^2p^2} - \frac{5\Omega'p^2}{2\Delta\Lambda p^3}\biggr]^{-1}. \label{RSEC}
\end{align}
\normalsize 
The RSE in Eq. \ref{RSEC} depends on the properties of the original graph $G$, i.e. $\Delta$, $\Lambda$, $\Phi$, $\Psi$, and $\Omega'$. Note that those properties are unknown for the third party. However, practitioners need to know the error bound of the estimator using the properties in the sample. To do so, we give the estimation of the variance as follows. 

Based on NES sampling scheme we have $\mathbb{E}(\Delta_g)=\frac{1}{3}\Delta p^2$, $\mathbb{E}(\Lambda_g)=\Lambda p$, $\mathbb{E}(\Phi_g)=\frac{8}{15}\Phi p^3$, $\mathbb{E}(\Psi_g)=\frac{2}{6}\Psi p$, and $\mathbb{E}(\Omega'_g)=\frac{5}{12}\Omega'p^2$. Thus, substitute $\Delta$, $\Lambda$, $\Phi$, $\Psi$, and $\Omega'$ in Eq. \ref{RSEC} by their estimations, and after some math work we obtain the estimator of the RSE as: 
	\begin{align}
	\widehat{RSE}(\widehat{\mathcal{C}})&\approx \biggl[\frac{1}{\Delta_g}+\frac{2\Phi_g}{\Delta^2_g}+ \frac{2\Psi_g}{\Lambda^2_g} - \frac{2\Omega'_g}{\Delta_g\Lambda_g }\biggr]^{-1}. \label{AproxRsecc1p}
	\end{align}
	where $\Phi_g$ is the number of pairs of dependent triangles, $\Psi_g$ is the number of pairs of shared wedges, and $\Omega'_g$ is the number of pairs of wedges and triangles with one common edge observed based on subgraph $g$.
The RSE in Eq. \ref{AproxRsecc1p} hangs on several variables in the sample, i.e. $\Delta_g$, $\Lambda_g$, $\Phi_g$, $\Psi_g$, and $\Omega'_g$. To have better understating the RSE of the estimator, we need to simplify it further more. We claim that when sampling probability $p$ is small, the first term in Eq. \ref{AproxRsecc1p}, i.e. $\Delta_g^{-1}$, is dominant. Therefore, we give the following Theorem as a simplified estimator for the RSE of $\mathcal{C}$.
\begin{theorem} \label{Theo:RSEapprox}
	When $p$ is small, the RSE of $\widehat{\mathcal{C}}$ is approximated by
	\begin{equation}
	\widehat{RSE}(\widehat{\mathcal{C}})\approx \Delta_g^{-1/2}. \label{approxRSEfirstterm}
	\end{equation}
\end{theorem}

\section{The bias-corrected estimator } \label{Sec:BiasCorr}
To quantify the bias, we apply the expectation on  $\widehat{\mathcal{C}}$: 
\begin{align}
\mathbb{E}\big(\widehat{\mathcal{C}}\;\big)=\mathbb{E}\biggl(\frac{3\Delta_g}{p\Lambda_g}\biggr)=\frac{3}{p}\mathbb{E}\biggl(\frac{\Delta_g}{\Lambda_g}\biggr).  \label{E_C1}
\end{align}
The approximation of $\mathbb{E}\big(\Delta_g/\Lambda_g\big)$ using the quadratic Taylor expansion of $\Delta_g/\Lambda_g$ in the neighborhood of $(a,b)$ is:
\begin{align}
\mathbb{E}\biggl(\frac{\Delta_g}{\Lambda_g}\biggr)\approx&\frac{a}{b}+\frac{a }{b^3}var(\Lambda_g)-\frac{1}{b^2}cov(\Delta_g,\Lambda_g). \label{E_Delta_g/Lambda_g}
\end{align}
Replace $a=\mathbb{E}\big(\Delta_g\big)$ and $b=\mathbb{E}\big(\Lambda_g\big)$ and take  $\mathbb{E}\big(\Delta_g\big)=\Delta p^2/3$ and $\mathbb{E}\big(\Lambda_g\big)=\Lambda p$; substitute Eq. \ref{E_Delta_g/Lambda_g} in Eq. \ref{E_C1}, we obtain:
\begin{align}
\mathbb{E}\big(\widehat{\mathcal{C}}\big)\approx&
\frac{\Delta}{\Lambda}\biggl(1+\frac{var(\Lambda_g)}{\Lambda^2p^2}-\frac{3cov(\Delta_g,\Lambda_g)}{\Delta\Lambda p^3}\biggr).
\end{align}
Note that $\mathcal{C}=\Delta/\Lambda$. Thus, we get:
\begin{align}
\mathbb{E}\big(\widehat{\mathcal{C}}\big)\approx&\mathcal{C}\biggl(1+\frac{var(\Lambda_g)}{\Lambda^2p^2}-\frac{3cov(\Delta_g,\Lambda_g)}{\Delta\Lambda p^3}\biggr). \label{E_C1p}
\end{align}
To quantify the bias of $\widehat{\mathcal{C}}$, the relative bias, i.e. RB= $(\mathbb{E}\big(\widehat{\mathcal{C}}\big)-\mathcal{C})/\mathcal{C}$, is used.  
By substituting Eq. \ref{E_C1p} in RB, and simplifying the result we obtained:

\begin{align}
RB\approx \frac{var(\Lambda_g)}{\Lambda^2p^2}-\frac{3cov(\Delta_g,\Lambda_g)}{\Delta\Lambda p^3}.	 \label{approx_RB_C_G}   
\end{align}

The bias depends on the variance of $\Lambda_g$ and the covariance between $\Delta_g$ and $\Lambda_g$. Using the same treatment in the variance of $\widehat{\mathcal{C}}$ for the $var(\Lambda_g)$ and $cov(\Delta_g,\Lambda_g)$, i.e. Eq. \ref{approxvarLambda} and Eq. \ref{approxCov},  and replacing them in Eq. \ref{approx_RB_C_G} we have

\begin{align}
RB\approx&\frac{2\Psi\;p}{3\Lambda^2p^2} - \frac{5\Omega'\;p^2}{4\Delta\Lambda p^3}.
\end{align}

Thus, we give our result in the following theorem.
\begin{theorem} \label{Theo_RB}
	The RB of the estimator is approximated by 
	\begin{align}
	RB\approx\frac{1}{p}\biggl[\frac{2\Psi}{3\Lambda^2} - \frac{5\Omega'}{4\Delta\Lambda }\biggr].	   
	\end{align}
	and its estimation is
	\begin{align}
	\widehat{RB}\approx \frac{2\Psi_g}{\Lambda_g^2} - \frac{\Omega'_g}{\Delta_g\Lambda_g }.	\label{AproxRBcc1p}   
	\end{align}
\end{theorem}
%
\begin{figure}[t]
	\centering
	\includegraphics[height=5.5cm, width=6.5cm,clip]{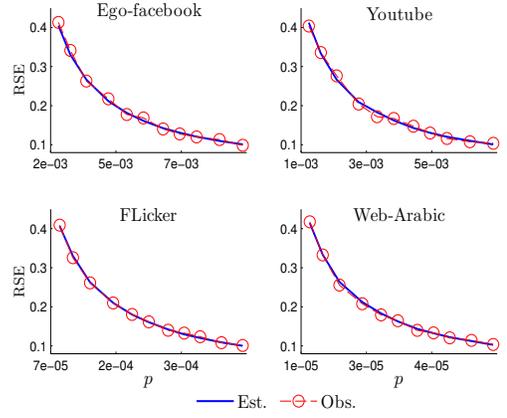}
	\caption{ The observed RSEs of $\widehat{\mathcal{C}}$ fit perfectly the estimated ones in representative graphs. Estimated RSEs are obtained based on Eq. \ref{AproxRsecc1p}. 
	}
	\label{Plot:RseCC}
\end{figure} 
Using $\widehat{RB}$, a bias-corrected estimator of $\widehat{\mathcal{C}}$ is  
\begin{align}
\widehat{\mathcal{C}}^+=\frac{\widehat{\mathcal{C}}}{1+\widehat{RB}}. 
\end{align}


%

\begin{table*}
	\caption{Properties of the networks in our experiments, sorted by graph size $N$.}
	\centering
	\scalebox{0.913}{		
		\begin{threeparttable}	
			{ 
				\setlength{\extrarowheight}{2pt}
				\begin{tabular}{rrrrrrrrrl}
					\toprule
					\multicolumn{1}{c}   {Network}& N$({\times 10^6})$ & M & $\mathcal{C}$ & $\Delta ({\scriptsize\times 10^6})$ &$\Lambda(\times 10^9)$& $\Phi(\times 10^9)$&$\Psi(\times 10^{12})$&$\Omega'(\times 10^{10})$& Description\\ \hline
					Ego-facebook\tnote{1} & 0.004& 88,234&0.519 & 4.8 &0.009&   0.2 & 0.003 & 0.1  & OSN \\
					CA-GrQc \tnote{1}&0.005 & 14,484 &0.629 & 0.1 & 0.0002& 0.002 & 0.00001 & 0.0009  &Collaboration  \\
					Wiki-vote\tnote{1} & 0.007&100,762 &0.125 & 1.8 & 0.014& 0.04 & 0.006 & 0.08 & OSN \\
					AstroPh \tnote{2} &0.01  &198,050  &0.31  &4.0  &0.012  & 0.07 & 0.002 & 0.07  & Citation \\
					CA-CondMat\tnote{1} & 0.02& 93,439 &0.264 & 0.5 & 2 &   0.002 & 0.0001 & 0.002&Coauthorship \\
					
					HepPh \tnote{2} &0.02  &3,148,447  &0.279  &587  &2  & 92  & 4  & 106 &  Coauthorship\\
					Enron-email\tnote{2} & 0.03 & 183,831 &0.085 & 2 &0.025& 0.03 & 0.01 & 0.09 &E-communication \\
					Brightkite\tnote{1}&0.05 & 214,078 &0.110 &1.4 &0.013& 0.02 & 0.005 & 0.03 &OSN \\	
					Facebook \tnote{2} &0.06  &817,035  &0.147  &10.5  &0.07  &0.1 & 0.02 & 0.2  & OSN \\
					Epinions \tnote{2} &0.07  &405,740  &0.065  &4.8  &0.07  & 0.1 & 0.06 & 0.3  & OSN \\
					
					Slashdot-Zoo \tnote{2} &0.07  &467,731  &0.023  &1.6  &0.06  & 0.02 & 0.05 & 0.09 & OSN \\
					Prosper \tnote{2} &0.08  &3,330,022  &0.003  &3.4  &1.1  & 0.06 & 1  & 0.5   &Interaction\\
					
					Livemocha \tnote{2} &0.1  &2,193,083  &0.014  &10.0 &0.716  &0.1 & 0.8 & 1 & OSN \\				
					Douban \tnote{2} &0.1  &327,162  &0.01  &0.1  &0.011  &  0.0001 & 0.001  & 0.001& OSN\\
					Gowalla \tnote{1}&0.1 & 950,327  & 0.023&6.8  & 0.290 & 0.1 & 2  & 0.7 &OSN \\				
					Libimseti \tnote{2} &0.2  &17,233,142   &0.007  &207  &28  & 19  & 262 & 151    &OSN\\
					Digg \tnote{2} &0.2  &1,548,126    &0.061 &42  &0.69  & 3  & 2 & 6  &  OSN\\
					
					Web-Stanford \tnote{2} &0.2  &1,992,636   &0.008  &33  &3.94  & 9  & 75  & 17   & Web graph\\
					Dblp-Coau\tnote{1} &0.3& 1,049,866    &0.306 & 6 &0.021 & 0.1 & 0.001 & 0.06 &Coauthorship \\
					Web-NotreDame\tnote{1} & 0.3 & 1,090,108   &0.087& 26  &0.304& 1.5 & 1 & 1 &Web graph \\
					Amazon\tnote{1}&0.3& 925,872   &0.205 & 2 &0.009& 0.003  & 0.0005 & 0.004&Co-purchasing  \\				
					Actor \tnote{2} &0.3  &15,038,083   &0.166  &1,040  &6.26  & 91  & 10  & 157   &  Collaboration\\
					
					Citeseer\tnote{2}&0.3 & 1,736,145   &0.049 & 4  &0.081& 0.01 & 0.02 & 0.06&Citation  \\
					Dogster \tnote{2}&0.4& 8,543,549  &0.014  & 250  &17& 42 & 378 & 191 &OSN \\ 
					Catster \tnote{2} &0.6  & 15,695,166   &0.028  &1,969  &69 & 1,017 & 3,651  & 1,528  & OSN \\
					Web-Berkeley \tnote{2} &0.6  & 6,649,470   &0.0069  &194 &27.9  & 105  & 1,148  & 176  & Web graph \\
					Web-Google \tnote{2}&0.8& 4,322,051   & 0.055 & 40  &0.727& 0.6 & 1  & 1 &Web graph \\ 
					
					Youtube\tnote{1}&1.1 & 2,987,624   &0.006 & 9  &1& 0.2 & 17  & 2 &OSN \\
					Dblp\tnote{2}&1.3& 5,362,414    &0.170 & 36  &0.214& 0.4 & 0.05 & 0.4 &Coauthorship \\				
					Hyves \tnote{2} &1.4  & 2,777,419    &0.001  &2  &1.4  & 0.02 & 32  & 0.1 &  OSN\\
					Wiki-Polish \tnote{2} &1.5  & 42,188,631    &0.01  &3,402  &308  & 697  & 62,290  & 3,466 & Web graph \\
					Trec-wt10g \tnote{2} &1.6  & 6,679,248   &0.014  &63  &4.3  & 11  & 50  & 13 &  Web graph\\
					
					Wiki-Japanese \tnote{2} &1.6  &56,231,610   &0.021  &3,863  &180  & 685 & 12,652 & 2,784 & Web graph \\
					Pokec \tnote{2} &1.6  &22,301,964   &0.046  &97   &2.08 & 0.7 & 3  & 2  & OSN\\
					As-skitter\tnote{1}&1.6 & 11,095,298   &0.005 & 86 &16& 20 & 291  & 48  &Internet topology \\
					
					Hudong \tnote{2} &1.9  &14,428,382    &0.003  &64  &18.7  & 5  & 357  & 20 & Web graph\\
					
					Hollywood \tnote{3} &1.9  &24,337,642  &0.152  &614  &4  & 27 & 2  & 37  & OSN \\
					Baidu \tnote{2} &2.1  &17,014,946   &0.002  &75  &30.8  & 4  & 1,717 & 69  & Web graph\\
					Flicker\tnote{2}&2.3& 22,838,276  &0.107  &  2,512 &23& 613 & 120  & 1,079 &OSN\\				
					Flixster \tnote{2} &2.5  &7,918,801    &0.013  &23  &1.7  & 0.3 & 1  & 1 &  OSN\\
					Wiki-Russian\tnote{2}  &2.8  &63,058,425    &0.015  &5,697 &370  & 1,180  & 56,397  & 5,420 & Web graph \\
					
					Wiki-Franch \tnote{2} &3.0  &83,455,052    &0.015  &6,843  &455  & 4,237 & 54,242 & 7,677 & Web graph \\
					Orkut\tnote{2}&3.0 & 117,185,083   &0.041 & 1,882 &45& 67 & 320  & 347 &OSN\\
					USpatent \tnote{2} &3.7  &16,518,947    &0.067  &22  &0.33  & 0.08 & 0.02  & 0.1  &  Citation\\
					LiveJournal\tnote{1}&3.9 & 34,681,189   &0.125 & 533 &4& 39 & 7 & 30  &OSN \\
					
					Web-Arabic\tnote{3}&22 & 553,903,073   & 0.031 & 110,686 &3,531& 112,260 & 986,071 & 91,940 &Web graph \\
					Twitter\tnote{2}&41 & 1,202,513,046   &0.0008 & 104,474 &123,435& 176,266 & 154,818,391 & 956,719 & OSN \\
					
					MicrosoftAc.G.\tnote{4}&46 & 528,463,861   &0.015 & 1,734 &115& 19 & 4,776 & 107 & Citation \\
					Friendster\tnote{2}&65 & 1,806,067,135   &0.017 & 12,521 &720& 185 & 1,138  & 1,284 &OSN \\
					\bottomrule
				\end{tabular}		
				\begin{tablenotes}\footnotesize
					\item[1] SNAP: \cite{snapnets}  \item [2] Konect: \cite{konect:2016} \item [3]\cite{BRSLLP,BoVWFI} \item [4] \cite{sinha2015overview}
				\end{tablenotes}	
			}
		\end{threeparttable}
	}
	\label{table_datasets1}
\end{table*}
 
The bias-corrected estimator depends on the structure of sampled graph $g$. Therefore, using a single pass over an edge stream of the input graph we need to count $\Delta_g$, $\Lambda_g$, $\Psi_g$, and $\Omega'_g$. Recall that $\Psi_g$ is the number of pairs of shared wedges and $\Omega'_g$ is the number of pairs of wedges and triangles with one common edge observed based on subgraph $g$. The authors in \cite{etemadi2017cbias} have shown that the cost of counting $\Psi_g$ and $\Omega'_g$ is the same as counting $\Lambda_g$ and $\Delta_g$. Note that the meaning of $\Omega'$ is different from the one in \cite{etemadi2017cbias}, i.e. $\Omega$ (the number of pairs of wedges and closed ones with a common edge) and here it is $\Omega'= (\Omega-6\Delta)/2$.

\begin{figure*}[ht!]
	\centering
	\includegraphics[height=11cm, width=16cm,clip]{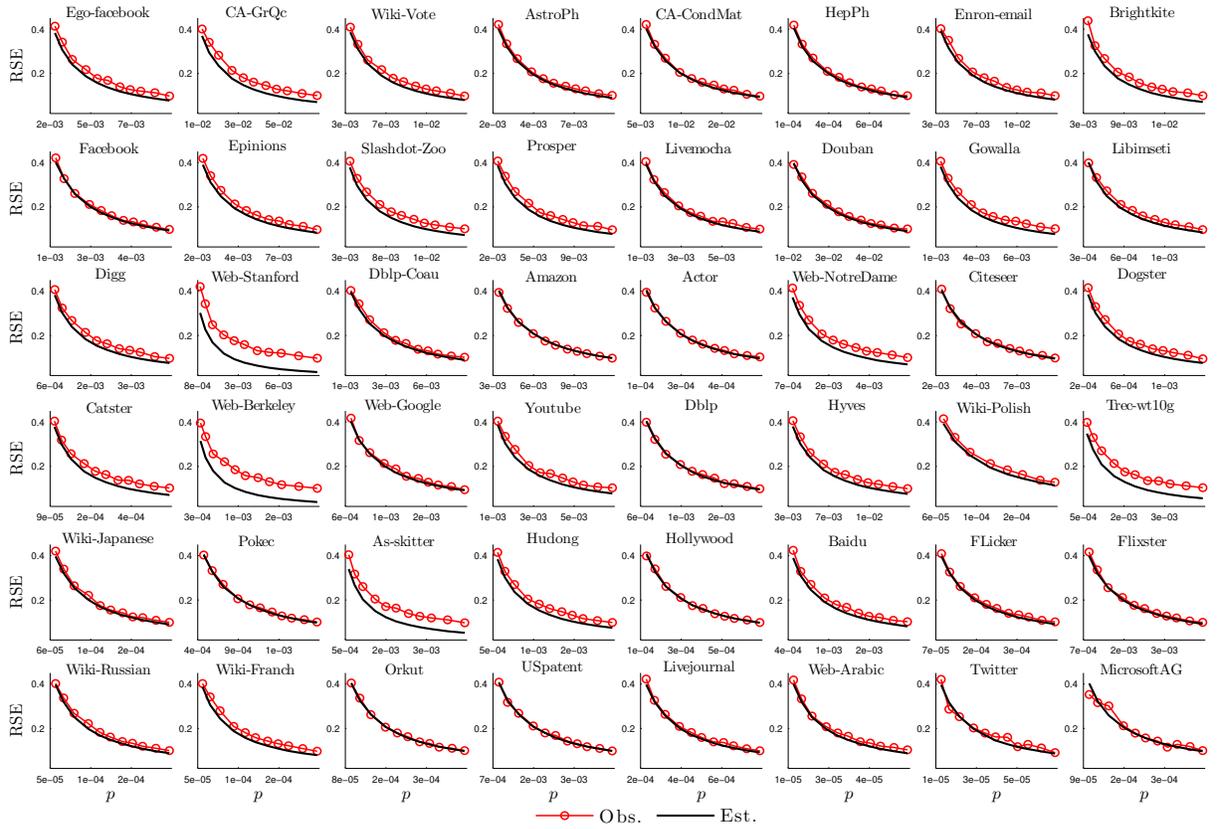}
	\caption{ The estimated RSEs obtained based on Eq. \ref{approxRSEfirstterm}  are apt estimations for the observed RSEs of $\widehat{\mathcal{C}}$.}
	\label{Plot:RseCCapprox}
\end{figure*}
\begin{figure}[t] 
	\centering
	\includegraphics[height=5cm, width=6.5cm,clip]{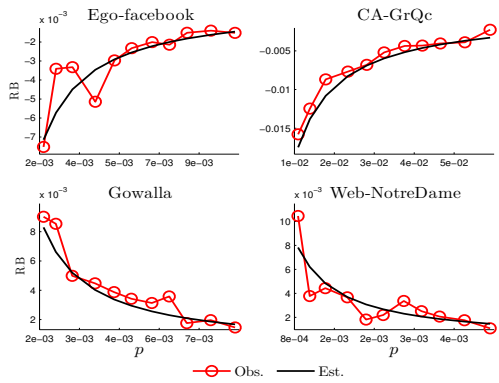}
	\caption{ The observed RBs of  $\widehat{\mathcal{C}}$ (biased estimator) support our estimations of RB based on Eq. \ref{AproxRBcc1p}. 
	}
	\label{Plotcc1pRB}
\end{figure}
\begin{figure}[t]
	\centering
	\includegraphics[height=5cm, width=6.5cm,clip]{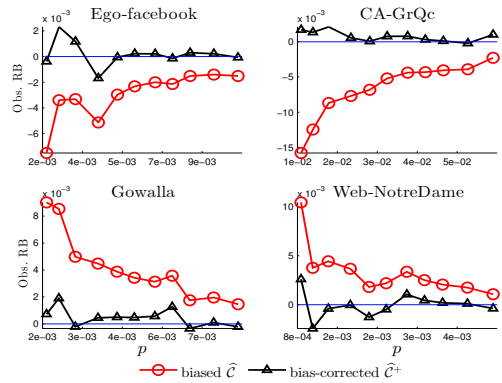}
	\caption{ Our biased-corrected $\widehat{\mathcal{C}}^+$ removes the bias perfectly. 
	}
	\label{Plotcc1pBiasCorrected}
\end{figure}

\section{Experiments}
\label{sec_Exp_Results}
The code along with all the data are publicly available at 
\url{myweb.cs.uwindsor.ca/~etemadir/cbiasstream.}
The observed RSEs and RBs are obtained over $k$ independent runs of the algorithm, where $k=1000$ except for the three largest networks with $k=100$ for RSE, and k=50,000 for RB. Carrying out the experiments, including obtaining the ground truth of the network properties,  is computationally expensive.  The experiments were accomplished  on two big servers each with 24 cores and 256 GB RAM. 


\begin{figure*}[t]
	\centering
	\includegraphics[height=5cm, width=17cm,clip]{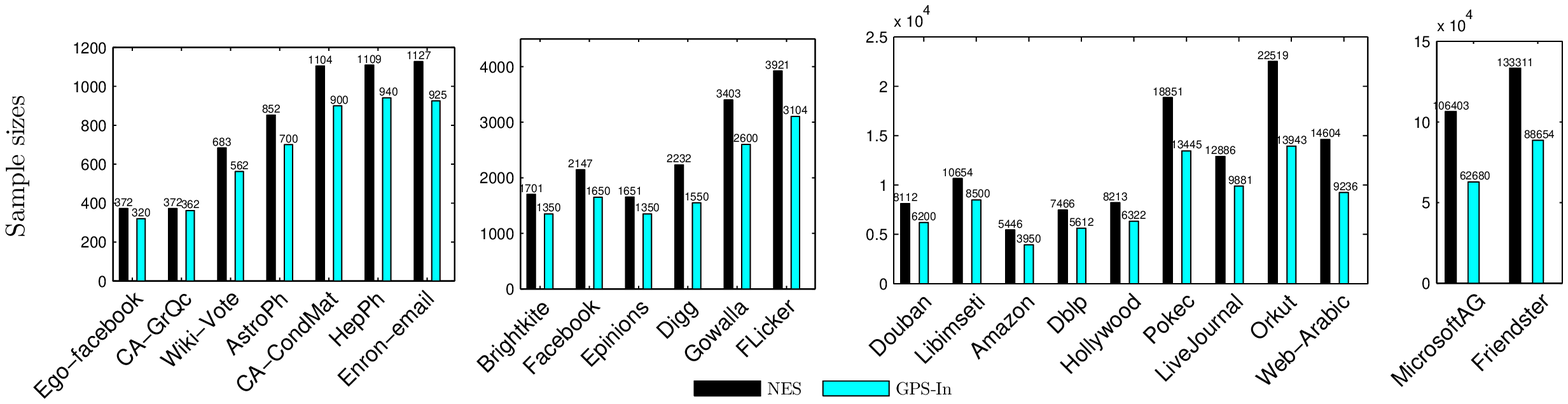}
	\caption{ Comparison of sample sizes, i.e. number of sampled edges, of the methods when RSE=0.2. The sample size of NES is comparable with GPS-In \cite{ahmed2017sampling}.}
	\label{PlotCompareWithGPSinMemory}
\end{figure*}

\subsection{Datasets} 
Our analytical results show that the variance and bias hangs on the structure of networks. Due to the fact that different networks have different structures, the bias and variance vary greatly from graph to graph. Furthermore, several approximations were made in the derivation of the theorems. To find out the patterns behind, we need to experiment extensively with graphs from variety of domains with different sizes and structures. We use 49 real network graphs in total. The networks are from several domains including online social networks (OSNs), web graph, co-authorship, citation, etc. The network sizes range from 14 thousand to 2 billion edges. For each network graph, self-loops were removed and directionally was ignored in digraphs. The statistics of the graphs are listed in Table \ref{table_datasets1}.  

\subsection{Verification of Theorem \ref{Theo:RSEapprox}}
 To verify the approximations made in the derivation of the RSE of the estimator in Theorem \ref{Theo:RSEapprox}, the parameters of the estimator were set up to obtain the RSEs between 0.1 and 0.4. First, we report the observed RSEs along with its estimations based on Eq. \ref{AproxRsecc1p} in Fig. \ref{Plot:RseCC}. Only 4 representative graphs are reported for lack of space. Similar pattern are observed for other graphs. We make several observations as follows.   
\begin{itemize}
	\item The estimated RSEs based on Eq. \ref{AproxRsecc1p} fit perfectly the observed RSEs not only for large graphs when $p$ is small but also for small ones. Take Ego-facbook, for example, the smallest graph in our datasets, the observed RSEs (red circles) fit very well our estimations based on Eq. \ref{AproxRsecc1p}. 
	\item When $p$ is small the first term in Eq. \ref{AproxRsecc1p}, i.e. $1/\Delta_g$, is a dominant term compared to the remaining three. 
	The reason is as follows. Firstly, when $p$ is small, the probability of identifying shared triangles, i.e. $\Phi_g$, and pairs of triangle and wedge with one common edge ($\Omega'_g$) based on the sampled $g$ is very small. Note that $\mathbb{E}(\Phi_g)=\frac{8}{15}\Phi p^3$, and $\mathbb{E}(\Omega'_g)=\frac{5}{12}\Omega' p^2$. Moreover, the term $2\Psi_g/\Lambda^2_g$ is ignorable due to the fact that their expectations depends on the sampling probability $p$ which it is in the same order in the both expectations. Therefore, in the sample $\Psi_g<<\Lambda_g^2$. 
	\item By increasing sampling probability $p$, terms $+2\Phi_g/\Delta^2_g$ and  $- 2\Omega'_g/\Delta_g\Lambda_g$ are increasing in the same order. Surprising that the two terms neutralize each other. Furthermore, term $2\Psi_g/\Lambda^2_g$ remains ignorable compared to the other terms.  
\end{itemize}
The observations above support our claim in Theorem \ref{Theo:RSEapprox} to simplify the RSE of the estimator. To verify the result in Theorem \ref{Theo:RSEapprox}, we report the observed RSEs and their estimations based on Eq. \ref{approxRSEfirstterm} in Fig. \ref{Plot:RseCCapprox}. It can be seen that the observed RSEs support our estimations not only for large graphs but also for small ones. In a few small graphs, i.e. Web-Standford, Web-Berkeley, and As-skitter, there are small gaps between the observed and estimated RSEs. However, by increasing the size of networks (see the last row in the figure) the estimated RSEs match perfectly the observed ones. Thus, we believe that our result in Theorem \ref{Theo:RSEapprox} can be used in practice to determine the size of samples to achieve a given accuracy level of the estimation. 

\subsection{The bias}
To understand the bias, we set the parameters of the estimator to achieve the RSEs between 0.1 and 0.4. The estimator run on the graphs and the observed RBs (relative bias) and the estimated ones based on Eq. \ref{AproxRBcc1p} were computed. The results were reported in Fig. \ref{Plotcc1pRB} for the graphs with the RB more than 0.7\%. We make several observations as follows: 
1) The observed RBs support the estimated RBs based on Eq. \ref{AproxRBcc1p} for all graphs;
2) Both negative and positive biases are observed;  
3) The largest RBs were observed on small graphs and it can be as high as 2\%;
4)  In most of the graphs in our datasets, the bias is very small and it is ignorable. 

We also report the observed RBs of our biased estimator and the bias-corrected one in Fig. \ref{Plotcc1pBiasCorrected}. It can be seen that the bias of $\widehat{\mathcal{C}}^+$ was removed in all the plots.  

\section{Discussions and Conclusions}
This paper addresses the estimation of the bias and variance in a streaming algorithm for estimating clustering coefficient. Essentially it is about estimating the properties of an estimator. It is important since it is the only way to know how good an estimate is during the estimation process. 

Our result is obtained based on two simplifications. One is the Taylor expansion--we take only the first two terms. The other is $1-p\approx1-p^2\approx 1$, assuming that sampling probability is very small if the data is very large. With such simplifications, we can characterize the variance with a single variable, i.e., $\Delta_g$. Although in theory variances depends on graph structures characterized by $\Phi$, $\Psi$, and $\Omega'$, all these variables can be neglected when estimations are performed on very large graph. This simple yet powerful result is very useful in practice-- we can give a confidence interval when an estimate is given. 

For the  bias part, we conclude that it is small overall, and it is more observable for smaller graphs. Interestingly, the bias can be either positive or negative, depending on the graph structure. 

Although our result is developed on NES algorithm, the same method can be extended to numerous other streaming algorithms.  Besides, NES itself is a very powerful algorithm. Despite its simplicity, its performance is comparable to the state-of-the-art algorithm GPS-In as illustrated in Fig.\ref{PlotCompareWithGPSinMemory}. Hence, the variance estimator and NES are a good combination to be used in practice. 

      
\section{Acknowledgments}
The research is supported by NSERC and CZI (Chan Zuckerberg Initiative).
\newcommand{\BIBdecl}{\setlength{\itemsep}{0.05 em}}
\bibliographystyle{IEEEtran}


\balance

\section*{Appendix} \label{Appen.A}
\begin{Lemma} \label{lem:varDelta}
	Let $\Delta_g$ be the number of closed wedges identified based on $g$ using NES. The variance of $\Delta_g$ is
	\begin{equation}
	var(\Delta_g)  
	= \frac{1}{3}\left(\Delta (p^2-\frac{1}{3}p^4) 
	+ 8\Phi (\frac{2}{5}p^3-\frac{1}{3}p^4)\right).
	\end{equation}
\proof 
Applying the variance we have
\begin{align}
var(\Delta_g) =var(\sum\limits_{i=1}^{\Delta} \delta_i)
=\sum\limits_{i=1}^{\Delta}\sum\limits_{j=1 }^{\Delta} cov( \delta_i, \delta_j) \notag \\
=\sum\limits_{i=1}^{\Delta} var( \delta_i) 
+ \sum\limits_{i \neq j} cov( \delta_i, \delta_j). \notag \label{VAR_CW_REG_main1}
\end{align}
We remind the reader that indicator $\delta_i$ is 1 if $i^{th}$ closed wedge is identified based on $g$; otherwise it is 0. 
The probability of identifying a closed wedge by NES is $\frac{1}{3}p^2$. Hence, the variance of $\delta_i$ is $\frac{1}{3}p^2-\frac{1}{9}p^4$. Therefore the cost of the variance term is $\Delta(\frac{1}{3}p^2-\frac{1}{9}p^4)$.  
The covariance between two independent closed wedges is zero. Thus, we need to find the covariance between dependent closed wedges. The probability of identifying such a dependent case is $\frac{2}{15}p^3$. Hence, the covariance between two shared closed wedges in such cases is $\frac{2}{15}p^3-\frac{1}{9}p^4$. Recall that the total number of pairs of shared triangles is denoted by $\Phi$. For each pair of shared triangles there are four dependent closed wedges. Thus, in summation above, there are $8\Phi$ pairs of shared closed wedges. Therefore, the cost of covariance term is $8\Phi(\frac{2}{15}p^3-\frac{1}{9}p^4)$. Add the costs of the two terms, we obtain the Lemma.
\end{Lemma}

\begin{Lemma} \label{lem:varLambda}
Suppose $\Lambda_g$ be the number of wedges identified based on $g$ using NES. The variance of $\Lambda_g$ is obtained by
	\begin{align}
	var(\Lambda_g)&= \Lambda (p-p^2) 
	+ \frac{2}{3}\Psi(p-p^2).
	\end{align}
	where $\Lambda$ and $\Psi$ are the number of wedges and the count of pairs of shared wedges in $G$ respectively.
\proof
Recall that $\lambda_i$ is an indicator for $i^{th}$ wedge in $G$.  Indicator $\lambda_i$ is 1  when $i^{th}$ wedge is identified based on $g$, and 0 else.  
The variance of $\Lambda_g$ is 
\begin{align}
var(\Lambda_g)	=var(\sum\limits_{i=1}^{\Lambda} \lambda_i)
=\sum\limits_{i=1}^{\Lambda}\sum\limits_{j=1 }^{\Lambda} cov( \lambda_i, \lambda_j) \notag \\
= \sum\limits_{i=1}^{\Lambda} var( \lambda_i) 
+ \sum\limits_{i \neq j} cov( \lambda_i, \lambda_j). \notag
\label{VAR_CW_}
\end{align}
 The probability of identifying a wedge by NES is $p$.  When $i=j$ holds, the covariance term is equal to the variance of $\lambda_i$, which is $p-p^2$. Thus, we get 
$var(\widehat{\Lambda})  =   \Lambda( p-p^2) + \sum\limits_{i \neq j} cov( w_i, w_j).$
Next, we need to understand the covariance between two wedges. When the two wedges $w_i$ and $w_j$ are independent the covariance between them is zero. Hence, we need to consider the covariance between dependent wedges. The probability of identifying a pair of dependent wedges is $p$. Thus, the covariance between two shared closed wedges is $(p-p^2)$. Suppose $\Psi$ be the number of pairs of dependent wedges in $G$. The chance to identify such a dependent case is $1/3$. Because $cov( \lambda_i, \lambda_j)=cov( \lambda_j, \lambda_i)$ , we need to multiply the covariance term by 2.  
Thus, the lemma is proved.
\end{Lemma} 

\begin{Lemma} \label{Lem:cov}
Covariance between $\Delta_g$ and $\Lambda_g$ is given by
	\begin{equation}
	cov(\Delta_g,\Lambda_g)  
	=   2\Delta (p^2-p^3) +\frac{5}{12}\Omega'(p^2-p^3).
	\end{equation}
	where $\Omega'$ is the number of pairs of wedges and triangles with one common edge.

\proof
Let indicators $\delta_i$ and $\lambda_j$ be the same as defined before.
The covariance is given by
\begin{equation}
cov(\Delta_g,\Lambda_g)  
= \sum\limits_{i=1}^{\Delta}\sum\limits_{j=1}^{\Lambda}cov(\delta_i,\lambda_j). \notag
\end{equation}
 
 Suppose $\Omega'$ be the exact number of pairs of wedges and triangles with one common edge in $G$.  The sampling probabilities of such a pair is $p^2$. Moreover, the chance to identify such a pair in the streaming model by NES is $5/12$. Therefore, the total cost is $\frac{5}{12}\Omega'(p^2-p^3)$. In addition, for each closed wedge of a triangle, there are two cases of pairs of wedges and closed ones with a common edge. The total cost for such cases is $6\Delta(\frac{1}{3}p^2+\frac{1}{3}p^3)$. Therefore, we get the lemma.

\end{Lemma} 



\end{document}